\documentclass[preprint,showpacs,amsmath,amssymb,aps,prd,nofootinbib]{revtex4}
\usepackage{amssymb,graphics,graphicx,epstopdf,color}

\def\gs{\gtrsim}
\def\ls{\lesssim}

\begin{document}
\title{\Large \bf Bimodal/Schizophrenic Neutrino As a Bridge between Inflation and Dark Energy}
\author{Chian-Shu Chen$^{1,3}$\footnote{chianshu@phys.sinica.edu.tw} and Chia-Min Lin$^{2}$\footnote{cmlin@phys.nthu.edu.tw}}
\affiliation{$^{1}$Physics Division, National Center for Theoretical Sciences, Hsinchu, Taiwan 300\\$^{2}$Department of Physics, National Tsing Hua University, Hsinchu, Taiwan 300\\$^3$Institute of Physics, Academia Sinica, Taipei, Taiwan 115}


\begin{abstract}
We consider the possibility to combine inflection point inflation and growing neutrino dark energy in a particular setup of bimodal/schizophrenic neutrino model in the framework of MSSM$\otimes U(1)_{B-L}$.  In order to obtain a varying neutrino mass, we make an assumption that a soft mass of $\tilde{Z'}_{BL}$ depends on a scalar field which plays the role of a cosmon field. We also investigate the possible effects of the cosmon field in the early universe such as modulated reheating and found that the induced density perturbation is negligible.

\end{abstract}
\maketitle

\section{Introduction}
Neutrino oscillations observed from the solar, atmospheric and laboratory experiments provide the mass squared differences of the three active neutrinos as $\Delta m^2_{12} = (7.59\pm0.20)\times10^{-5}\rm{eV^2}$, $|\Delta m^2_{23}|=(2.43\pm0.13)\times10^{-3}\rm{eV^2}$, and mixing angles $\sin^2(2\theta_{12}) = 0.87\pm0.03$, $\sin^2(2\theta_{23}) > 0.92$, and $\sin^2(2\theta_{13}) < 0.15$\cite{pdg}. The firm evidences indicate the neutrinos have nonzero but tiny masses and at least one of the neutrino masses is of order $10^{-2}$ eV. Many scenarios have been proposed to explain the small neutrino masses. One of the most appealing solutions is introducing the right-handed neutrinos $N_{R}$ to the standard model (SM), in which $N_{R}$ may form the Dirac masses with left-handed neutrinos only or acquire Majorana masses. Different patterns of neutrino mass matrices can be casted and agree with the experimental data. One of the possibilities has been pointed out recently is the so-called bimodal/schizophrenic neutrinos~\cite{Allahverdi:2010us}, whereas the neutrino mass eigenstates can be part Dirac and part Majorana. Unlike the conventional pseudo-Dirac case~\cite{pseudo-Dirac} where the Majorana mass entry has to be less than at least $10^{-9}$ eV to satisfy the current solar data~\cite{deGouvea:2009fp}, the Majorana and Dirac mass matrix elements are all comparable for bimodal/schizophrenic neutrino scenario. Interestingly this consideration can be naturally realized in the framework of the minimal supersymmetric standard model (MSSM) with additional gauge $U(1)_{B-L}$ symmetry~\cite{Mohapatra:1986aw, Mohapatra:1986bd, Barry:2010en}, here B and L stand for baryon and lepton number respectively. The additional gauge anomaly from the $B-L$ symmetry can be cancelled by introducing the right-handed neutrino to each generation, which are singlets under the SM gauge group. This symmetry also sheds light on the fate of R-parity symmetry in MSSM~\cite{Mohapatra:1986su, Font:1989ai, Martin:1992mq}. Recent works~\cite{FileviezPerez:2008sx, Barger:2008wn} show that the breaking of $B-L$ and R-parity symmetries can be made by utilizing the right-handed sneutrino VEVs, and two light sterile neutrinos~\cite{Ghosh:2010hy, Barger:2010iv} are predicted in this model.

Furthermore an inflaton field $\phi$ defined along the D-flat direction $\phi = \frac{\tilde{N}_{R} + H_{u} + \tilde{L}}{\sqrt{3}}$ under MSSM$\otimes U(1)_{B-L}$~\cite{Allahverdi:2007wt}, in which the relevant superpotential is
\begin{eqnarray}\label{susypotential}
W = W_{\rm{MSSM}} + Y_{\nu}LH_{u}N_{R},
\end{eqnarray}
where $L$ and $H_{u}$ are the superfields denoting left-handed leptons and the Higgs doubelt which gives the up-type fermion masses respectively. It shows that a small Yuakwa coupling $Y_{\nu}$ of order $10^{-12}$ can naturally explain the neutrino mass which can be implemented in the scheme of bimodal/schizophrenic neutrino for Dirac type~\cite{Allahverdi:2010us} and also satisfies the observed density perturbations $P_{\zeta}^{1/2} \sim 5\times10^{-5}$ via  the so-called inflection point inflation~\cite{Allahverdi:2007wt} which combines the soft mass and $A$-term for $\phi$ field along the minimal angular direction. Therefore the small neutrino masses and inflationary universe are intimately connected to each other.

On the other hand, different cosmological observations~\cite{pdg} suggest the expansion of our universe has been accelerating recently since  about redshift of $z\simeq0.5$. The origin of the dominant component of energy density, dark energy, with negative pressure which drives the accelerated expansion of the universe is unknown. Models to explain dark energy include the cosmological constant~\cite{Weinberg:1988cp}, a dynamical scalar field called "quintessence"~\cite{Wetterich:1987fm, Ratra:1987rm} rolling in a very flat potential, or the modified gravity~\cite{Dvali:2000rv}. It was proposed that a "growing matter" or an associated component with neutrinos~\cite{Amendola:2007yx, Wetterich:2007kr} may be the mechanism to stop a dynamical evolution of the cosmon $\chi$, the dark energy scalar field once the neutrinos become non-relativistic and trigger the accelerated expansion of the universe. The neutrino masses are, therefore, depend on the time evolution of the cosmon. In this work we associate the Majorana component of bimodal/schizophrenic neutrino with the cosmon field in the framework of MSSM$\otimes U(1)_{B-L}$. Hence this Majorana  component is treated as the growing neutrino whose mass is related to the soft mass $\tilde{Z}'_{BL}$ and, of course, the $B-L$ breaking scale. We assume the supersymmetry breaking scale is associated with the field value of the cosmon, and therefore, evolve with time.

This paper is organized as follows. In section 2, we briefly review the model of MSSM$\otimes U(1)_{B-L}$ and its neutrino spectrum, especially the bimodal/schizophrenic active neutrino. In section 3 we discuss the inflection point inflation as a result of the flat direction $LH_{u}N_{R}$, here the smallness Yukawa of Dirac neutrino component play the role. In section 4 we take the Majorana part as the growing neutrino whose mass depends on the field value of cosmon $\chi$ to study the dark energy. In section 5, we consider the possible effects of the cosmon field in the early universe. Section 6 is our conclusion.

\section{The Framework and Neutrino Masses}

We consider the minimal supersymmetric extension of the standard model with additional $U(1)_{B-L}$ gauge group. The field content is the same of MSSM plus three additional right-handed neutrinos $N_{R_{i}} (i= 1,2,3)$, one for each generation, carrying $B-L$ charge $1$ to cancel the $B-L$ anomaly. The representations of the superfields under $SU(3)_{C}\times SU(2)_{L}\times U(1)_{Y}\times U(1)_{B-L}$ are given by
\begin{eqnarray}
Q &=&\left(\begin{array}{c}u \\d\end{array}\right) \sim (3,2,1/6,1/3), \quad u^c \sim (3,1,-2/3,-1/3), \quad d^c \sim (3,1,1/3,-1/3), \nonumber \\
L &=& \left(\begin{array}{c}\nu \\e\end{array}\right) \sim (1,2,-1/2,-1), \quad e^c \sim (1,1,1,1), \quad N_{R} \sim (1,1,0,1), \nonumber \\
H_{u} &\sim& (1,2,1/2,0), \quad {\rm and} \quad H_{d} \sim (1,2,-1/2,0).
\end{eqnarray}
Here we omit the family indices and $u (\nu)$ and $d (e)$ represent up- and down-type quarks (leptons) respectively.  Note that unlike the conventional seesaw mechanism, the right-handed Majorana masses $M_{R}$ is prohibited by the $U(1)_{B-L}$. One way to generate $M_{R}$ is through the breaking of $B-L$ symmetry by the vacuum expectation value (VEV) of the right-handed sneutrino $\langle \tilde{N}_{R}\rangle$~\cite{Mohapatra:1986aw, FileviezPerez:2008sx, Barger:2008wn, Hayashi:1984rd}. The point is that one can always rotate to the VEVs of right-handed sneutrinos to one direction. Let's take $\langle \tilde{N}_{R3}\rangle \neq 0$ and $\langle \tilde{N}_{R1}\rangle = \langle \tilde{N}_{R2}\rangle = 0$, the neutrino mass matrix is
\begin{eqnarray}\label{numass}
m_{\nu} = \left(\begin{array}{ccc}0_{3\times3} & m_{D} & -g_{BL}\langle\tilde{\nu}_{L}\rangle^T \\m^T_{D} & 0_{3\times3} & M^T_{BL} \\-g_{BL}\langle\tilde{\nu}_{L}\rangle & M_{BL} & M_{\rm SUSY}\end{array}\right)
\end{eqnarray}
in the basis of $(\nu_{Li}, N_{Ri}, \tilde{Z'}_{BL})$, where $M_{BL} = (0, 0, g_{BL}\langle\tilde{N}_{R3}\rangle)$ and $m_{D} = Y_{\nu}\langle H_{u}\rangle$. $g_{BL}$ is the gauge coupling of $U(1)_{B-L}$ and $M_{\rm SUSY}$ is the supersymmetry (SUSY) breaking Majorana mass of $\tilde{Z}_{BL}$. We will neglect the contributions of left-handed sneutrino VEVs, which mix the left-handed neutrinos with neutralinos and are bounded $\langle\tilde{\nu}_{L}\rangle\ls 1$ MeV if $M_{\rm SUSY} \approx 1$ TeV from the neutrino data. Since the breaking scale of $B-L$ symmetry, $\langle\tilde{N}_{R3}\rangle$, is higher than the electroweak scale $\langle H_{u,d}\rangle = v_{u,d}$, we first generate one heavy right-handed Majorana neutrino from the lower right $2\times2$ block matrix which reads
\begin{eqnarray}\label{heavyMajorana}
M_{N_{R3},\tilde{Z'}_{BL}} = \left(\begin{array}{cc}0 & g_{BL}\langle\tilde{N}_{R3}\rangle \\g_{BL}\langle\tilde{N}_{R3}\rangle & M_{\rm SUSY}\end{array}\right).
\end{eqnarray}
From Eqs.~(\ref{numass}) and~(\ref{heavyMajorana}) one obtains one heavy and two light sterile neutrino with mass eigenvalues $M_{R3}$ and $M_{R1,2}$ respectively~\cite{Mohapatra:1986aw, Ghosh:2010hy, Barger:2010iv}. This leads to the possibility that the three light active neutrinos are bimodal/schizophrenic form~\cite{Barry:2010en}: one linear combination get a Majorana mass, denoted $m_{\nu_{3}} \simeq \frac{Y^2_{\nu_{3}}v_{u}^2}{M_{R3}}$, and two combinations of Dirac masses are given by $m_{\nu_{1,2}} \simeq  Y_{\nu_{1,2}}v_{u}$. The effective Yukawa interactions of neutrino masses can be
written as
\begin{eqnarray}
L_{\nu} = Y_{\nu_{1}}L_{1}H_{u}N_{R1} + Y_{\nu_{2}}L_{2}H_{u}N_{R2} + \frac{Y^2_{\nu_{3}}}{M_{R3}}(L_{3}H_{u})^2 + {\rm H.c.}.
\end{eqnarray}
The idea of bimodal/shizophrenic neutrino can be tested by measuring the flux ratios of high energy neutrinos from extragalaxies and the effective mass of neutrinoless double beta decay~\cite{Barry:2010en}. 
We will argue that the small Yukawa's of Dirac component is suitable for the so-called inflection point inflation and the Majorana part can be treated as growing matter to explain the dark energy in the following sections.

\section{Inflection Point Inflation}
In the following discussion, we set the reduced Planck mass $M_P=1$ for simplicity.
The scalar potential along the D-flat direction $\phi = \frac{\tilde{N}_{R} + H_{u} + \tilde{L}}{\sqrt{3}}$ (with the superpotential $W \sim h \phi^3$) is given by~\cite{Allahverdi:2007wt}
\begin{eqnarray}
V(\phi) &=&\frac{1}{2}m^2 \phi^2 - AW + \big|\frac{\partial W}{\partial \phi}\big|^2  \nonumber \\
        &=&\frac{1}{2}m^2 \phi^2 - \frac{A h}{6\sqrt{3}}\phi^3+\frac{h^2}{12}\phi^4,
\end{eqnarray}
where $m$ is the SUSY breaking soft mass, $AW$ the SUSY breaking $A$ term, and $A=4m$ is assumed in order to obtain a saddle point $\phi=\phi_0=\sqrt{3}m/h$ where we have $V^{\prime}(\phi_0)=V^{\prime\prime}(\phi_0)=0$. A phase is also been fixed in order to produce a negative A-term. Near the saddle point $V(\phi_0)=m^4/4h^2$, we can expand the potential as $V(\phi)=V(\phi_0)+(1/3!)V^{\prime\prime\prime}(\phi_0)(\phi-\phi_0)^3$, where $V^{\prime\prime\prime}=2hm/\sqrt{3}$. Here the potential is so flat that we can have inflation with a much lower scale (lower than, say, GUT scale). Like many small field inflation models, inflation ends when we have the slow roll parameter $|\eta| \sim 1$, this means:
\begin{equation}
|\eta| \equiv \big|\frac{V^{\prime\prime}}{V}\big|=2 \left( \frac{2h}{m} \right)^2 \big|\frac{(\phi-\phi_0)}{\phi_0}\big| \sim 1,
\end{equation}
where $\phi$ should be recognized as the field value at the end of inflation $\phi_{end}$. The number of e-folds is given by
\begin{equation}
N=\int^{\phi}_{\phi_{end}}\frac{V}{V^{\prime}}d\phi \sim \left( \frac{m}{2h} \right)^2 \frac{\phi_0}{(\phi_0-\phi)}.
\end{equation}
Please note that we have $\eta \sim -2/N$. This implies the spectral index is given by
\begin{equation}
n_s \sim 1+ 2\eta = 1-\frac{4}{N}.
\end{equation}
For the scale of Cosmic Microwave Background (CMB), that is $N=50$, we have $n_s=0.92$. The constraint for the parameters $m$ and $h$ is from the CMB temperature fluctuation. The spectrum is given by
\begin{equation}
P_{\zeta}=\frac{1}{12\pi^2}\left( \frac{V}{V^{\prime}} \right)^2 V=\frac{h^4}{9\pi^2 m^2}N^4.
\end{equation}
The constraint imposed by CMB temperature fluctuation ($\Delta T/ T \sim 10^{-5}$) is
\begin{equation}
P^{1/2}_{\zeta}=\frac{h^2}{3\pi m}N^2=5 \times 10^{-5}.
\end{equation}
This means if the soft mass $m$ is of the TeV scale, and with $N \sim 50$, we need $h \sim 10^{-12}$. This is also the right value for a neutrino Dirac mass term.
\section{Growing Neutrino}

In order to have dark energy, we assume there is a cosmon field $\chi$ which has an exponential form of the potential therefore result in a ``scaling solution"~\cite{Wetterich:1987fm}. This means the energy density of the cosmon field is a constant fraction of the background contents (i.e. radiation or matter).
In the framework of SUSY, the potential can be obtained in the following way. Consider a superpotential $W=\Lambda^{3+\gamma} Q^{-\gamma}$ and a K\"ahler potential $K=-\ln(Q+Q^\ast)$~\cite{Brax:1999gp, Copeland:2000vh}. This kind of form is present for the dilaton and moduli fields in string theory. The scalar potential in supergravity is given by (in terms of the canonically normalized field $\chi \equiv \ln Q/\sqrt{2}$):
\begin{eqnarray}
V &=& e^K[(W_i+WK_i)K^{j^\ast i}(W_j+WK_j)^\ast-3|W|^2] \nonumber  \\
  &=& M^4 e^{-\alpha \chi}
\end{eqnarray}
where the subscript $i$ indicates the derivative with respect to the $i$-th field and $M^4=\Lambda^{5+\kappa}(\kappa^2-3)/2$ with $\kappa \equiv 2\gamma+1$ and $\alpha \equiv \sqrt{2}\kappa$. The scaling solution means the fraction of dark energy density is a constant depends on $\alpha$
\begin{equation}
\label{fraction}
\Omega_h=\frac{n}{\alpha^2},
\end{equation}
with $n=3(4)$ for the matter (radiation) dominated epoch. We will discuss the possible role of $\chi$ during inflation epoch in the next section. 

Here we assume that the soft mass $M_{\rm{SUSY}}$ of $\tilde{Z'}_{BL}$ is determined by the expectation value of $Q$ which we parameterized as $M_{\rm{SUSY}} =\overline{M} Q^{-\epsilon/\sqrt{2}}$. In terms of $\chi$, we have $M_{\rm{SUSY}} = \overline{M} \exp(-\epsilon \chi)$. After diagonalization of Eq.~(\ref{heavyMajorana}), we obtain\footnote{Here we assume $M_{BL} > M_{\rm{SUSY}}$ and keep the expansion of $M_{\rm{SUSY}}/M_{BL}$ to first order.}
\begin{equation}
M_{N_{R3}} \sim M_{BL} \left[ 1-\frac{1}{\tau} \exp(-\epsilon \chi) \right],
\end{equation}
here $\tau = M_{BL}/\overline{M}$ and $\chi_t=-\ln \tau/\epsilon$ is defined as $M_{N_{R3}}(\chi_t)=0$. Notice that the dependence on cosmon field is the same form as suggested in ref.~\cite{Wetterich:2007kr}. Therefore we can get the neutrino mass through seesaw mechanism
\begin{equation}
m_\nu=\overline{m}_{\nu} \{ 1-\exp[-\epsilon (\chi-\chi_t)] \}^{-1}
\end{equation}
with $\overline{m}_{\nu} = \frac{Y^2_{\nu3}v^2_{u}}{M_{BL}}$. For $\epsilon < 0$, the neutrino mass increases with increasing $\chi$. The equation of motion of $\chi$ is
\begin{eqnarray}
\label{cosmon}
\ddot{\chi}+3H\dot{\chi}&=&\frac{\partial V}{\partial \chi}+\beta(\chi)(\rho_\nu-3p_\nu), \nonumber \\
\beta(\chi)&=&-\frac{\partial}{\partial \chi} \ln m_\nu(\chi)=\frac{1}{\chi-\chi_t}.
\end{eqnarray}
Here $\rho_\nu$ and $p_\nu$ are the neutrino energy density and pressure, obeying 
\begin{equation}
\label{neutrino}
\dot{\rho}+3H(\rho_\nu+p_\nu)=-\frac{\dot{\chi}}{\chi-\chi_t}(\rho_\nu-3p_\nu).
\end{equation}
For $\chi$ near $\chi_t$ we have
\begin{equation}
m_\nu(\chi)=\frac{\beta(\chi)}{\epsilon}\overline{m}_{\nu}.
\end{equation}
When $\chi$ approaches $\chi_t$, $\beta$ becomes very large and stops the evolution of $\chi$ and $V(\chi_t)$ would behave like a cosmological constant which plays the role of dark energy. If $M \sim 1$, for $\alpha \chi_t \sim 276$ the cosmological constant has a value compatible with observation. This amounts to the condition $\epsilon=-\alpha \ln \tau/276$.
From Eq.~(\ref{fraction}), upper bounds on early dark energy require $\alpha \gs 10$~\cite{Doran:2007ep}. We choose $\alpha=10$ and $\chi_t=27.6$. Therefore for $\ln \tau={\cal O}(1)$, we have $\epsilon \sim -0.05$. This means a rather mild $\chi$-dependence of the soft mass.

In order to determine the current $\beta(t_0)$, we have to solve Eqs.~(\ref{cosmon}) and (\ref{neutrino}) numerically by requiring $\Omega_m(t_0)=0.24$~\cite{Wetterich:2007kr}. By using the above parameters, the result is $\beta(t_0)=314$. Therefore if the present neutrino mass $m_\nu(t_0)=0.44$eV, we have $\overline{m}_{\nu}=7 \times 10^{-5}$eV. That means if we take the TeV seesaw, $M_{N_{R3}}\sim {\cal O}(1)$ TeV, the scale $M_{BL}$ is around $10^{3}$ TeV and Yukawa coupling $Y_{\nu3} \sim {\cal O}(10^{-6})$ for $v_{u} = 100$ GeV.

\section{Cosmon in the Early Universe}
It is interesting to consider the possible roles played by the cosmon field in the early universe. The potential for the cosmon field $\chi$ is given by
\begin{equation}
V=M^4 e^{-\alpha \chi}
\end{equation}
where we assume $\alpha=10$ and $M=1$ therefore
\begin{equation}
V^{\prime\prime}=100 e^{-10\chi}.
\end{equation}
We note here that due to the large $\alpha$ the cosmon field cannot be the inflaton field because if it dominates the energy density of our universe we have the slow roll parameter
\begin{equation}
\eta=\frac{V^{\prime\prime}}{V}=100>1.
\end{equation}
We cannot choose a small $\alpha$ because of Eq.~(\ref{fraction}). In our scenario, during inflation we have the Hubble parameter $H^2 \sim (m^2/h)^2 \sim 10^{-36} \sim V(\phi)$. Compare $V^{\prime\prime}$ with $H^2$, we found that when $\chi>8.7$, the cosmon field becomes slow-rolling. At this time its potential is $V(\chi) < 10^{-38}$ which is subdominant. Inflation normally lasts for $N \gg 50$, therefore the cosmon potential is expected to be even smaller hence observable inflation (start around $N=50$) is not affected.

However there maybe some interesting cosmological consequences due to the quantum fluctuation of the cosmon field during inflation. For example, the inflaton field mainly decays into $\tilde{Z'}_{BL}$ whose mass depends on the expectation value of the cosmon field as $M_{\tilde{Z'}_{BL}} \sim M_{BL} \left[ 1 + \frac{1}{\tau} \exp(-\epsilon \chi) \right]$. Since we assume the neutrino is Dirac type we ignore its mass in final state. The decay width of the inflaton depends on the mass of $\tilde{Z'}_{BL}$ through
\begin{eqnarray}
\Gamma = \frac{g^2_{BL}m_{\tilde{N}_{R}}}{8\pi}\big(1 - \frac{m^2_{\tilde{Z}'_{BL}}}{m^2_{\tilde{N}_{R}}}\big)^2,
\end{eqnarray}
so the quantum fluctuation of the cosmon field $\delta \chi \sim H/2\pi$ at $N=50$ may result in a contribution to the CMB temperature fluctuation $\delta \Gamma/ \Gamma \sim \Delta T/T$. This effect is called ``inhomogeneous preheating (or modulated reheating)"~\cite{Kofman:2003nx}. However, the dependence of the decay width on the mass is weak in our scenario so the effects is actually negligible. Even though this shows that it is a nontrivial task when we try to combine two scenarios with one for inflation and the other for dark energy.

\section{Conclusion}
We have shown that bimodal/schizophrenic neutrino model can accommodate both inflation and dark energy in the framework of MSSM$\otimes U_{B-L}(1)$, in which the SM three active neutrinos can divide into two Dirac and one Majorana components. The inflection point inflation can be realized along the flat direction of the Dirac type interactions. And we assume that the soft mass of $\tilde{Z}'_{BL}$ depends on the cosmon field $\chi$. This implies the Majorana component of active neutrinos behaves as the growing matter, and hence the $\chi$ field becomes the dark energy in current universe. 
 
We have also shown that the cosmological consequence of the quantum fluctuation of the cosmon field is negligible here. However, had we choose a different inflation scenario the effect may be strong enough and may even produce detectable non-Gaussianity. If so, this could imply the seeds of structure formation and CMB temperature fluctuation is actually from dark energy. We will investigate this possibility in our future work.

\section*{Acknowledgement}
CML was supported by the
NSC under grant No. NSC 99-2811-M-007-068 and CSC was supported by the National Center of Theoretical Sciences of Taiwan (NCTS).


\end{document}